\documentclass[twoside]{article}
\usepackage{graphics}
\usepackage{latexsym}

\catcode`\@=11
\long\def\@makefntext#1{ \protect\noindent \hbox
to 3.2pt {\hskip-.9pt
$^{{\eightrm\@thefnmark}}$\hfil}#1\hfill}       

\def\@makefnmark{\hbox to 0pt{$^{\@thefnmark}$\hss}}    

\def\ps@myheadings{\let\@mkboth\@gobbletwo
\def\@oddhead{\hbox{}
\rightmark\hfil\eightrm\thepage}
\def\@oddfoot{}\def\@evenhead{\eightrm\thepage\hfil
\leftmark\hbox{}}\def\@evenfoot{}
\def\sectionmark##1{}\def\subsectionmark##1{}}

\oddsidemargin=\evensidemargin 
\newcounter{sectionc}\newcounter{subsectionc}\newcounter{subsubsectionc}
\renewcommand{\section}[1] {\vspace{12pt}\addtocounter{sectionc}{1}
\setcounter{subsectionc}{0}\setcounter{subsubsectionc}{0}\noindent
    {\tenbf\thesectionc. #1}\par\vspace{5pt}}
\renewcommand{\subsection}[1] {\vspace{12pt}\addtocounter{subsectionc}{1}
\setcounter{subsubsectionc}{0}\noindent
{\bf\thesectionc.\thesubsectionc. {\kern1pt \bfit
#1}}\par\vspace{5pt}}
\renewcommand{\subsubsection}[1] {\vspace{12pt}\addtocounter{subsubsectionc}{1}
    \noindent{\tenrm\thesectionc.\thesubsectionc.\thesubsubsectionc.
    {\kern1pt \tenit #1}}\par\vspace{5pt}}

\newcounter{appendixc}
\newcounter{subappendixc}[appendixc]
\newcounter{subsubappendixc}[subappendixc]
\renewcommand{\thesubappendixc}{\Alph{appendixc}.\arabic{subappendixc}}
\renewcommand{\thesubsubappendixc}
    {\Alph{appendixc}.\arabic{subappendixc}.\arabic{subsubappendixc}}

\renewcommand{\appendix}[1] {\vspace{12pt}
        \refstepcounter{appendixc}
        \setcounter{figure}{0}
        \setcounter{table}{0}
        \setcounter{lemma}{0}
        \setcounter{theorem}{0}
        \setcounter{corollary}{0}
        \setcounter{definition}{0}
        \setcounter{equation}{0}
        \renewcommand{\thefigure}{\Alph{appendixc}.\arabic{figure}}
        \renewcommand{\thetable}{\Alph{appendixc}.\arabic{table}}
        \renewcommand{\theappendixc}{\Alph{appendixc}}
        \renewcommand{\thelemma}{\Alph{appendixc}.\arabic{lemma}}
        \renewcommand{\thetheorem}{\Alph{appendixc}.\arabic{theorem}}
        \renewcommand{\thedefinition}{\Alph{appendixc}.\arabic{definition}}
        \renewcommand{\thecorollary}{\Alph{appendixc}.\arabic{corollary}}
        \renewcommand{\theequation}{\Alph{appendixc}.\arabic{equation}}
        \noindent{\tenbf Appendix \theappendixc #1}\par\vspace{5pt}}
\newcommand{\subappendix}[1] {\vspace{12pt}
        \refstepcounter{subappendixc}
        \noindent{\bf Appendix \thesubappendixc. {\kern1pt \bfit #1}}
    \par\vspace{5pt}}
\newcommand{\subsubappendix}[1] {\vspace{12pt}
        \refstepcounter{subsubappendixc}
        \noindent{\rm Appendix \thesubsubappendixc. {\kern1pt \tenit #1}}
    \par\vspace{5pt}}

\newcommand{\textlineskip}{\baselineskip=13pt}
\newcommand{\smalllineskip}{\baselineskip=10pt}

\newcommand{\copyrightheading}[1]
    {\vspace*{-2.5cm}\smalllineskip{\flushleft
    {\footnotesize International Journal of Modern Physics D, #1}\\
    {\footnotesize \copyright\kern2pt World Scientific Publishing
     Company}\\
     }}

\newcommand{\publisher}[2]{{\begin{center}\footnotesize\smalllineskip
    Received #1\\
    Revised #2
    \end{center}
    }}

\def\abstracts#1#2#3{{
    \centering{\begin{minipage}{4.5in}\footnotesize\baselineskip=10pt
    \parindent=0pt #1\par
    \parindent=15pt #2\par
    \parindent=15pt #3
    \end{minipage}}\par}}

\newcommand{\bibit}{\nineit}

\renewenvironment{thebibliography}[1]
        {\frenchspacing
     \ninerm\baselineskip=11pt
         \begin{list}{\arabic{enumi}.}
        {\usecounter{enumi}\setlength{\parsep}{0pt}
     \setlength{\leftmargin 12.7pt}{\rightmargin 0pt}
         \setlength{\itemsep}{0pt} \settowidth
    {\labelwidth}{#1.}\sloppy}}{\end{list}}

\newcounter{itemlistc}
\newcounter{romanlistc}
\newcounter{alphlistc}
\newcounter{arabiclistc}
\newenvironment{itemlist}
        {\setcounter{itemlistc}{0}
     \begin{list}{$\bullet$}
    {\usecounter{itemlistc}
     \setlength{\parsep}{0pt}
     \setlength{\itemsep}{0pt}}}{\end{list}}

\newcommand{\fcaption}[1]{
        \refstepcounter{figure}
        \setbox\@tempboxa = \hbox{\footnotesize Fig.~\thefigure. #1}
        \ifdim \wd\@tempboxa > 5in
           {\begin{center}
        \parbox{5in}{\footnotesize\smalllineskip Fig.~\thefigure. #1}
            \end{center}}
        \else
             {\begin{center}
             {\footnotesize Fig.~\thefigure. #1}
              \end{center}}
        \fi}

\newcommand{\tcaption}[1]{
        \refstepcounter{table}
        \setbox\@tempboxa = \hbox{\footnotesize Table~\thetable. #1}
        \ifdim \wd\@tempboxa > 5in
           {\begin{center}
        \parbox{5in}{\footnotesize\smalllineskip Table~\thetable. #1}
            \end{center}}
        \else
             {\begin{center}
             {\footnotesize Table~\thetable. #1}
              \end{center}}
        \fi}

\def\@citex[#1]#2{\if@filesw\immediate\write\@auxout
    {\string\citation{#2}}\fi
\def\@citea{}\@cite{\@for\@citeb:=#2\do
    {\@citea\def\@citea{,}\@ifundefined
    {b@\@citeb}{{\bf ?}\@warning
    {Citation `\@citeb' on page \thepage \space undefined}}
    {\csname b@\@citeb\endcsname}}}{#1}}

\newif\if@cghi
\def\cite{\@cghitrue\@ifnextchar [{\@tempswatrue
    \@citex}{\@tempswafalse\@citex[]}}
\def\citelow{\@cghifalse\@ifnextchar [{\@tempswatrue
    \@citex}{\@tempswafalse\@citex[]}}
\def\@cite#1#2{{$\null^{#1}$\if@tempswa\typeout
    {IJCGA warning: optional citation argument
    ignored: `#2'} \fi}}

\def\pmb#1{\setbox0=\hbox{#1}
    \kern-.025em\copy0\kern-\wd0
    \kern.05em\copy0\kern-\wd0
    \kern-.025em\raise.0433em\box0}

\def\fnt#1#2{\footnotetext{\kern-.3em
    {$^{\mbox{\scriptsize #1}}$}{#2}}}

\def\fpage#1{\begingroup
\voffset=.3in
\thispagestyle{empty}\begin{table}[b]\centerline{\footnotesize #1}
    \end{table}\endgroup}

\def\runninghead#1#2{\pagestyle{myheadings}
\markboth{{\protect\footnotesize\it{\quad #1}}\hfill}
{\hfill{\protect\footnotesize\it{#2\quad}}}} \headsep=15pt

\font\tenrm=cmr10 \font\tenit=cmti10 \font\tenbf=cmbx10
\font\bfit=cmbxti10 at 10pt \font\ninerm=cmr9 \font\nineit=cmti9
 \font\eightrm=cmr8

\def\qed{\hbox{${\vcenter{\vbox{              
   \hrule height 0.4pt\hbox{\vrule width 0.4pt height 6pt
   \kern5pt\vrule width 0.4pt}\hrule height 0.4pt}}}$}}

\begin{document}
\setlength{\textheight}{7.7truein}    
\runninghead{Structure and properties of neutron stars in the Relativistic Mean-Field Theory
 $\ldots$} {Structure and properties of neutron stars in the Relativistic Mean-Field Theory $\ldots$}

\normalsize\textlineskip
\thispagestyle{empty}
\setcounter{page}{1}

\copyrightheading{}     

\vspace*{0.88truein}
\fpage{1}
\centerline{\bf STRUCTURE AND PROPERTIES OF NEUTRON STARS}
\centerline{\bf  IN THE RELATIVISTIC
                MEAN-FIELD THEORY}

\vspace*{0.035truein}
\vspace*{0.37truein}
\centerline{\footnotesize
             ILONA BEDNAREK\footnote{bednarek@us.edu.pl} \
         and RYSZARD MANKA\footnote{manka@us.edu.pl}}
\vspace*{0.015truein}
\centerline{\footnotesize\it Institute of
Physics, University of Silesia, Uniwersytecka 4,
 Katowice 40-007, Poland.}
\baselineskip=10pt
\vspace*{0.225truein}
\publisher{(received date)}{(revised date)}
\vspace*{0.21truein}
\abstracts{Properties of rotating neutron stars with the use of
relativistic mean-field theory are considered. The performed
analysis of neutron star matter is based on the nonlinear
Lgrangian density. The presence of nonlinear  interaction of vector mesons
modifies the density dependence of the $\rho$ field and
influences bulk parameters of neutron stars. The observed
quasi-periodic X-ray oscillations of low mass X-ray binaries can
be used in order to constrain the equation of state of neutron
star matter. Having assumed that the maximum frequency of
the quasi periodic oscillations originates at the circular orbit
it is possible to estimate  masses and radii of neutron stars. }{}{}
\vspace*{1pt}\textlineskip
\section{Introduction}
\vspace*{-0.5pt} \noindent Computation of the  equation of
state is one of the key problems in the construction of a
reliable model of a neutron star. Therefore the aim of this paper
is to study such neutron star parameters, namely the masses and
radii, which are the most sensitive ones to the form of the equation of
state. The influence of rotation on these parameters is also
estimated. The properties of neutron star matter in high-density
and neutron-rich regime are considered with the use of the
relativistic mean-field approximation. This approach implies the
interaction of nucleons through the exchange of meson fields, so
the model considered here comprises: nucleons, electrons and
scalar, vector-isoscalar and vector-isovector mesons. Consequently
the contributions coming from these components  characterize
the pressure and energy density. In this paper
the nonlinear vector-isoscalar  self-interaction is dealt. Modification
of this type was proposed by Bodmer  in order to achieve good
agreement with the Dirac-Br\"{u}ckner \cite{1} calculations at
high densities. The first version of the $\rho$ meson field
introduction s of a minimal type without any nonlinearity and
consists only of the coupling of this field to  nucleons.  This
case is enlarged by the nonlinear vector-isoscalar and
vector-isovector interaction which modifies the density
dependence of the $\rho$ mean field and the energy symmetry. Such an
extension of the neutron star model was inspired by the paper
\cite{2} in which the authors indicate that there exists a
relationship between the neutron-rich skin of a heavy nucleus and
the properties of a neutron star crust. For a more realistic
neutron star description the composite equation of state has been
employed. The presented forms of the EOS for neutron rich matter
have been combined with those of Negele and Vautherin \cite{3}, the
Machleidt-Holinde-Elster Bonn potential \cite{4}
  and
Haensel-Pichon \cite{5}. It has been found that neutron stars of
minimal configurations strongly depend on the proton fraction
which in turn is connected with the presence of the nonlinear
$\omega -\rho $ interaction.
The observed X-ray radiation can be used to investigate phenomena occuring in the strong gravitational fields near neutron stars.
The Rossi X-Ray Timing Explorer discovered high-frequency X-ray brightness oscillations from several low-mass
X-ray binaries.
Observations of kilohertz QPO sources can show the existence of circular orbits around neutron stars.
Using the relation between the orbit radius $r$ and the mass of
the neutron star for given value of the observed frequency $\nu _{QPO}$
one can see that the orbit radius is determined by the mass of a neutron star which in turn depends on
the particular choice of the equation of state.

The outline of this paper is as follows.
In Sect.1 the employed
equations of state are introduced with and without the additional
$\omega -\rho$ meson interaction. Then the impact of the introduction of
this nonlinear interaction on the neutron star parameters is
studied.
In
Sect.2  general properties of nonrotating and rotating neutron
stars are calculated. The results are employed to construct
the solutions of the equations of motion obtained for a
particle confined to the  equatorial plane. The obtained results for the chosen form of the EOS
followed by the discussion on their implications are
presented in Sect3.

\section{The equation of state}
\vspace*{-0.5pt}
\noindent
It is the equation of state (EOS) of the matter which determines the characteristic of a neutron star.
Thus the reliable form of the equation of state  is the basic input to the structure equations.
However, the knowledge of the EOS at densities around and beyond nuclear density $(\rho_{0} \simeq 2.5 \times 10^{14} g/cm^3)$
is incomplete.
In general the properties of nuclear matter at extreme conditions
remain one of the most important problem. There have been several attempts
to determine the proper form of
the equation of state for dense nuclear matter.
The first one is the nonrelativistic Bre\"{u}ckner-Bethe theory based
on the use of the free nucleon-nucleon interaction with a
variational method for the many body correlations. An
alternative approach is the relativistic field -theoretical method.
The effective relativistic field theory of
interacting hadrons makes an optional approach to nuclear many
body problem and makes the description of
the bulk properties of finite nuclei and
binding energy of nuclear matter reliable and reproduces the
experimental data
\cite{6,7}.
These theories are effective ones, as coupling constants are
determined by the bulk properties of nuclear matter such as
saturation density, binding energy, compressibility and the
symmetry energy see also \cite{8}. The bulk properties put certain limits on these
theories and enable extrapolation to higher densities.
The goal
is to extend the RMF theory to the system with density reaching the level of
several values of the nuclear density.
However, the complete and more realistic description of a neutron star requires taking into consideration not only the interior region
of a neutron star but also the remaining layers, namely the inner and outer crust and the surface.
The adequate density regions are presented below:
\begin{itemlist}
\item 2$\times 10^3<\rho <1\times 10^{11}$-light metals, electron gas
\item $1\times 10^{11}< \rho < 2\times 10^{13}$-heavy metals, relativistic electron gas
\item $2\times 10^{13}<\rho <5\times 10^{15}$- nucleons, relativistic leptons
\end{itemlist}
In this paper the composite equation of state for the entire neutron star density span was constructed by
joining together the equation of state of the neutron rich matter region given by equations
(\ref{en:dens})(\ref{ps:ress}), the Negele-Vautherin
EOS  and Bonn
for
the relevant density range  between $10^{14}$ and $5\times 10^{10} g/cm^3$ and the Haensel-Pichon EOS  for the density region
$9.6\times 10^{10} g/cm^3$ to $3.3\times 10^{7} g/cm^3$.
Since the density drops steeply near the surface of a neutron star, these layers do not contribute
to the total mass of a neutron star in a significant manner.
The inner neutron rich region up to density $\rho \sim 10^{13} g/cm^3$ influences decisively the neutron star structure and evolution.
Different forms of the equations of state are presented in
Fig.1. There are  original NL1 and TM1 parameter sets
describing the neutron rich star interior and the composite
EOS constructed by adding  Bonn and Negele-Vauterin equations of
state to the TM1 one. Of particular interest
is the influence of
the proton fraction $Y_{p}=n_{p}/n_{B}$ ($n_{p}$ and $n_{B}$ are the proton and baryon number densities respectively) on the behaviour of the EOS. There are
two cases presented in this figure with distinctively different proton fractions, namely the $Y_{p}=0.07$
and $Y_{p}=0.17$.
In Fig.2 the form of the equation of state for the original NL3 and NL3 parameter sets and for the NL3 with the additional $\Lambda _{V}$
coupling equals 0.025.
The simplest but unrealistic neutron star description assumes the presence of
neutrons only but
it is not possible for neutron star matter to be purely neutron
one.
The model presented in our work comprises baryons interacting
through the exchange of scalar and vector mesons, which gives the
Lagrange density function ${\cal L}_{0}$ consisting of the parts describing the free baryon and meson
fields together with the interaction terms.
In this model we are dealing with the electrically neutral neutron
star matter being in \( \beta  \)-equilibrium.
Therefore the imposed constrains, namely the charge neutrality and
$\beta $-equilibrium,
imply the presence of leptons. Mathematically it is expressed by adding the Lagrangian of free relativistic leptons
${\cal L}_{L}=\overline{\psi}_{e}(i\gamma ^{\mu }\partial _{\mu
}-m_{e})\psi_{e}$
to the Lagrangian function ${\cal L}_{RMF}$. Neutrinos are neglected here since they leak out from
the neutron star, whose energy diminishes at the same time.
Such a matter possesses a highly asymmetric character  caused
by the presence of small amounts of protons and electrons.
\begin{eqnarray}
\label{lagrang}
{\cal L}_{RMF} & = & \frac{1}{2}\partial _{\mu }\varphi \partial ^{\mu
}\varphi -\frac{1}{4}R_{\mu \nu }^{a}R^{a\mu \nu
}-\frac{1}{4}\Omega_{\mu \nu }\Omega^{\mu \nu
}+\frac{1}{2}M^{2}_{\omega }\omega _{\mu }\omega ^{\mu} \\ \nonumber
&+&\frac{1}{2}M^{2}_{\rho }\rho ^{a}_{\mu }\rho ^{a\mu }
+  g^2_{\rho}\rho ^{a}_{\mu }\rho ^{a\mu }\Lambda _{V}g^2_{\omega}\omega_{\mu}\omega^{\mu} - U(\varphi )\\
\nonumber
            & + & \frac{1}{4!}\xi (\omega _{\mu }\omega ^{\mu })^{2}+i\overline{\psi }\gamma ^{\mu }D_{\mu }\psi
-\overline{\psi }(M-g_{s}\varphi )\psi
\end{eqnarray}
The field tensors $R_{\mu \nu }^a$ and $\Omega _{\mu \nu }$ and the
covariant derivative $D_{\mu}$ are given by
\begin{equation}
R_{\mu \nu }^{a}=\partial _{\mu }\rho ^{a}_{\nu }-\partial _{\nu }\rho ^{a}_{\mu } +
g_{\rho}\varepsilon_{abc}\rho _{\mu }^{b}\rho _{\nu }^{c}
\end{equation}
\begin{equation}
\Omega_{\mu \nu}=\partial _{\mu }\omega _{\nu }-\partial _{\nu }\omega _{\mu }
\end{equation}
\begin{equation}
D_{\mu}=\partial_{\mu}+ ig_{\omega}\omega_{\mu}+\frac{1}{2}ig_{\rho}\rho_{\mu}^{a}\sigma^{a}
\end{equation}
The potential function \( U(\varphi ) \) possesses a polynomial form
introduced by Boguta and Bodmer \cite{9} in order to get a correct value of the
compressibility $K$ of nuclear matter at saturation density.
\begin{equation}
U(\varphi )=\frac{1}{2}m^{2}_{s}\varphi ^{2}+\frac{1}{3}g_{2}\varphi ^{3}+\frac{1}{4}g_{3}\varphi ^{4}
\end{equation}
The nucleon mass is
denoted by \( M \) whereas \( m_{s} \), \( M_{\omega } \), \(
M_{\rho } \) are masses assigned to the meson fields. The
parameters entering the Lagrangian function (\ref{lagrang}) are the coupling
constants \( g_{\omega } \), \( g_{\rho } \) and \( g_{s} \) and
the self-interacting coupling constants \( g_{2} \) and \( g_{3}
\). The Lagrangian function also includes the nonlinear term \(
\frac{1}{4}\xi g^{4}_{\omega }(\omega _{\mu }\omega ^{\mu })^{2}
\) which affects the form of the equation of state. This
modification was proposed by Bodmer in order to achieve good
agreement with the Dirac-Br\"{u}ckner calculations at high
densities.
This model is additionally supplemented by a new nonlinear $\omega-\rho$
coupling which causes causes the change in the density
dependence
of the $\rho$ field.
The parameters employed in this model are collected in
Table \ref{tab1}. It contains the TM1, NL1, NL3 and NL3s parameter
sets.
The ground state of a neutron star is thought to
be the question of  equilibrium dependence on the baryon
and electric-charge conservation. Neutrons are  main
components of a neutron star when the density of the matter is
comparable to the nuclear density. Both the proton and electron
numbers, which should be the same, are determined by the
equilibrium with respect to the reaction
\begin{equation}
n\leftrightarrow p+e^{-}+\nu _{e}
\end{equation}
After the leak out of neutrinos, the equilibrium of  neutron star matter is
established by the lowest energy states of fermions satisfying the following
relation
\begin{equation}
\mu _{p}=\mu _{n}-\mu _{e}
\end{equation}
\begin{table}[htbp]
\tcaption{\label{tab1} Parameter sets employed in this paper.}
\centerline{\footnotesize\smalllineskip
\begin{tabular}{l c c c c}
\hline
Parameter&
 NL1\cite{6}&
 NL3\cite{6}&
 NL3s\cite{10}&
 TM1\cite{11}\\
\hline
\( M \)&
 \( 938\, MeV \)&
 \( 938\, MeV \)&
 \( 938\, MeV \)&
 \( 938\, MeV \)\\
\hline
\( M_{w} \)&
 \( 795.4\, MeV \)&
 \( 782.5\, MeV \)&
 \( 795.359\, MeV \)&
 \( 783\, MeV \)\\
\hline
\( M_{\rho } \)&
 \( 763\, MeV \)&
 \( 763\, MeV \)&
 \( 763\, MeV \)&
 \( 770\, MeV \)\\
\hline
\( m_{s} \)&
 \( 492\, MeV \)&
 \( 508.2\, MeV \)&
 \( 492\, MeV \)&
 \( 511.2\, MeV \)\\
\hline
\( g_{2}\)&
 \( 12.17\, fm^{-1} \)&
 \( 2.03\, fm^{-1} \)&
 \( 12.17\, fm^{-1} \)&
 \( 7.23\, fm^{-1} \)\\
\hline
\( g_{3} \)&
 \( 0 \)&
 \( 1.666 \)&
 \( -36.259 \)&
 \( 0.618 \)\\
\hline
\( g_{s} \)&
 \( 10.077 \)&
 \( 9.696 \)&
 \( 10.138 \)&
 \( 10.029 \)\\
\hline
\( g_{\omega } \)&
 \( 13.866 \)&
 \( 12.889 \)&
 \( 13.285 \)&
 \( 12.614 \)\\
\hline
\( g_{\rho } \)&
 \( 8.488 \)&
 \( 8.544 \)&
 \( 9.264 \)&
 \( 9.264 \)\\
\hline
\( g_{\omega }^{4}\xi /6 \)&
 \( 0 \)&
 \( 0 \)&
 \( 0 \)&
 \( 71.308 \) \\
\hline
\(\Lambda_{V}\)&
 \( 0 \)&
 \( 0 \)&
 \( 0.025 \)&
 \( 0 \) \\
\hline
\end{tabular}}
\end{table}
The field equations derived from the Euler-Lagrange equations for meson fields
\( \varphi  \), \( \omega _{\mu } \) and \( \rho _{\mu }^{a} \) are the Klein-Gordon
equations with source terms coming from the baryon fields. They
are coupled differential equations
\begin{equation}
\label{egg1}
\partial _{\mu }\partial ^{\mu }\, \varphi +m_{s}^{2}\varphi +g_{2}\varphi ^{2}+g_{3}\varphi ^{3}=g_{s}\overline{\psi }\psi
\end{equation}
\begin{equation}
\label{egg2}
\partial _{\mu }\Omega^{\mu \nu }+M^{2}_{\omega }\omega ^{\nu }+\frac{\xi}{6} g_{\omega }^{4}(\omega _{\nu }\omega ^{\nu })\omega ^{\mu }+2g_{\rho}^2g_{\omega}^2\Lambda_{V}\omega^{\nu}\rho_{\mu}^a\rho^{\mu a}=g_{\omega }\overline{\psi }\gamma ^{\nu }\psi
\end{equation}
\begin{equation}
\label{egg3}
D_{\mu }R^{\mu \nu a}+M^{2}_{\rho }\, \rho ^{\nu a}+ 2g_{\rho}^2g_{\omega}^2\Lambda_{V}\omega_{\mu}\omega^{\mu}\rho_{\nu}^a=g_{\rho }\overline{\psi }\gamma ^{\nu }\sigma ^{a}\psi .
\end{equation}
Sources that appear in the equations of motion  are the baryon
current \[J_{B}^{\nu}=(\rho_{B},\overline{J}_{B}^{a})=\overline{\psi }\gamma ^{\nu }\psi \]
and  the isospin
current which exists only in the asymmetric matter
\[J^{\nu}=(\rho ,\overline{J})=\frac{1}{2}\overline{\psi}\gamma ^{\nu}\sigma ^{a}\psi .\]
In this model we are dealing with static, homogenous, infinite
matter, which sets certain simplifications on the Euler-Lagrange
equations, namely the derivative terms in the equations for meson
fields vanish due to translational invariance of infinite matter,
the rotational symmetry causes the disappearance of  spatial
components of vector meson fields. In the mean field approach the
meson fields are treated as classical ones after replacing them by
their mean value ($\varphi_{0}=<\varphi>, \omega _{0}=<\omega ^0>, \rho_{03}=<\rho ^{03}>$  ), baryon currants appearing in the Euler-Lagrange
equations are replaced by their ground state expectation value
which is the quantum selfconsistent fermion system.
In the case of the $\rho$ meson  only the neutral state is kept.
Having inserted the stated above simplifications the field
equations are reduced to
\begin{equation}
m_{s}^{2}\varphi _{0} +g_{2}\varphi_{0}^{2}+g_{3}\varphi _{0}^{3}=g_{s}\rho _{s}
\end{equation}
\begin{equation}
M_{\omega }^{2}\omega _{0}+\xi g_{\omega }^{4}\omega_{0}^{3}+(g_{\omega}g_{\rho})^2\Lambda_{V}\omega_{0}\rho_{03}^2=g_{\omega
}J^{0} _{B}
\end{equation}
\begin{equation}
M_{\rho }^{2}\rho_{03}+(g_{\omega}g_{\rho})^2\Lambda_{V}\omega_{0}^2\rho_{03}=\frac{1}{2}g_{\rho
}J^{0}_{3} \label{isovect}
\end{equation}
The densities occurring in these equations are the scalar density
$\rho _{s}$ and the conserved  baryon density $\rho _{B}$ and the isospin
density $\rho _{3}$. They are defined as
\begin{equation}
\rho _{s}=<\overline{\psi }\psi > =\frac{4}{(2\pi )^3}\int_{0}^{k_F}\frac{d^3kM^{*}}{\sqrt{k^2+M^{*2}}}
\end{equation}
\begin{equation}
\rho _{B}=<\overline{\psi }\gamma ^{0}\psi > = \rho _{p}+\rho _{n}
\end{equation}
\begin{equation}
\rho _{3}=<\overline{\psi }\gamma ^{0}\sigma ^{3}\psi > = \frac{1}{2}(\rho _{p}-\rho
_{n})
\end{equation}
where $\rho _{p}$ and $\rho _{n}$ are the proton and neutron
densities.
The expression describing the scalar density \( \rho _{s} \) relates the value of
the Fermi momentum $k_F$ and the baryon density \( \rho _{B} \).\\
The Dirac equations for  baryons that are obtained from the Lagrangian
function (\ref{lagrang}) have the following form
\begin{equation}
i\gamma ^{\mu }D_{\mu }\psi -(M-g_{s}\varphi _{0})\psi =0
\end{equation}
with $M^{*}=M-g_{s}\varphi _{0}$ being the effective nucleon mass which is generated by the
nucleon and scalar field interaction.
The isovector  meson field $\rho $ is considered two-fold
in this paper. The first is that of a  meson field introduced in
a minimal version and we are dealing with the simplest possible
coupling of the $\rho$ meson to  nucleons. There is no
mutual interaction between the
$\rho$ and $\omega$ mesons, which meets the requirements $\Lambda
_{V}=0$ and the equation (\ref{isovect}) gives simplfied density
dependence for the field $\rho _{03}$
\begin{equation}
\rho_{03}=\frac{g_{\rho}}{2M^2_{\rho}}(n_p-n_n).
\end{equation}
Taking into account the additional $\omega $- $\rho $ interaction
(the coupling constant $\Lambda_{V}\neq 0$) the self-consistency
equations for vector mesons were obtained.
The mutual
interactions of vector meson fields  result in the
effective meson masses which can be obtained with the help of the
following substitution in the Euler-Lagrange equations
\begin{equation}
M^2_{eff,\omega }
=M_{\omega}^2+2g^2_{\omega}g^2_{\rho}\Lambda_{V}\rho_{03}^2
\end{equation}
and
\begin{equation}
M^2_{eff,\rho}
=M_{\rho}^2+2g^2_{\omega}g^2_{\rho}\Lambda_{V}\omega^2.
\end{equation}
The forms of the effective masses of mesons versus the baryon number density $n_{B}$ are presented in Fig.3.
The symmetry energy $E_s$ dependence on the isovector density takes the
following form
\begin{equation}
E_s=\frac{g^2_{\rho}}{2M^2_{\rho}}(n_p-n_n)^2+3\Lambda_{V}(g_{\rho}g_{\omega})^2\rho^2_{03}\omega^2_{0}.
\end{equation}
Fig.4 shows the dependence of $\rho$ field expectation value on the baryon number density $\rho_{B}$ for the following  parameter sets:
the original NL3 ($\Lambda_{V}=0$) and for the extreme values of the parameter $\Lambda_{V}$.
Due to the growth of the meson effective mass the value of the $\rho$ field is significantly smaller in the presence of the $\omega -\rho $ coupling.
The proton fraction $Y_{p}$ versus the baryon number density $n_{B}$ is presented in Fig.5.
The pressure and the energy density are related to the trace of the energy
momentum tensor $T_{\mu \nu}$: \( P=\frac{1}{3}<T_{ii}> \), \( \epsilon =<T_{00}>
\)
\begin{equation}
T_{\mu \nu}=2\frac{\partial {\cal L}}{\partial g^{\mu \nu}}-g_{\mu \nu}{\cal L}
\end{equation}
where the Lagrangian function ${\cal L}={\cal L}_{RMF}+{\cal L}_{L}$. The total pressure \( P \) of the
neutron star matter consists of pressures coming from the
fermion and meson fields
\begin{equation}
P=\frac{1}{2}M_{\rho }^{2}\rho_{03}^{2}+\frac{1}{2}M_{\omega}^{2}\omega_{0}^{2}+\frac{1}{24}g_{\omega }^{4}\xi \omega_{0}^{4}+g^2_{\rho}g^2_{\omega}\Lambda_{V}(\omega^2_{0}+\rho^2_{03})-U(\varphi_{0})+P_{F}
\label{ps:ress}
\end{equation}
with the
fermion pressure \( P_{F} \) being the sum of the lepton
and nucleon contributions
\begin{equation}
P_{F}=\sum_{p,n}\frac{1}{3\pi^2}\int_{0}^{k_{F}}\frac{k^4dk}{\sqrt{k^2+(M-g_s\varphi_{0})^2}}+\frac{1}{3\pi^2}\int_{0}^{k_F}\frac{k^4dk}{\sqrt{k^2+m^2_e}}
\end{equation}
Analogously, the total energy
density \( \varepsilon  \) includes terms coming from meson and fermion fields
\begin{eqnarray}
\varepsilon & = & \frac{1}{2}g_{\rho }\rho_{3}-\frac{1}{2}M_{\rho}^2\rho_{03}^2+g_{\omega}\rho_{B}-g^2_{\rho}g^2_{\omega}\Lambda_{V}(\omega^2_{0}  + \rho^2_{03}) \\ \nonumber
& - & \frac{1}{2}M_{\omega }^{2}\omega_{0}^{2}-\frac{1}{24}g_{\omega }^{4}\xi \omega_{0}^{4} +U(\varphi _{0} )+\epsilon _{f}
\label{en:dens}
\end{eqnarray}
with $\varepsilon _{F}$
\begin{equation}
\varepsilon_{F}=\sum_{p,n}\frac{1}{3\pi^2}\int_{0}^{k_{F}}k^2dk\sqrt{k^2+(M-g_s\varphi_{0})^2}+\frac{1}{3\pi^2}\int_{0}^{k_F}k^2dk\sqrt{k^2+m^2_e}.
\end{equation}
Fig.6 shows
the pressure $P$ as a function of the energy density \( \varepsilon  \) for different parameter sets.
\begin{figure}
{\par\centering \resizebox*{10cm}{!}{\includegraphics{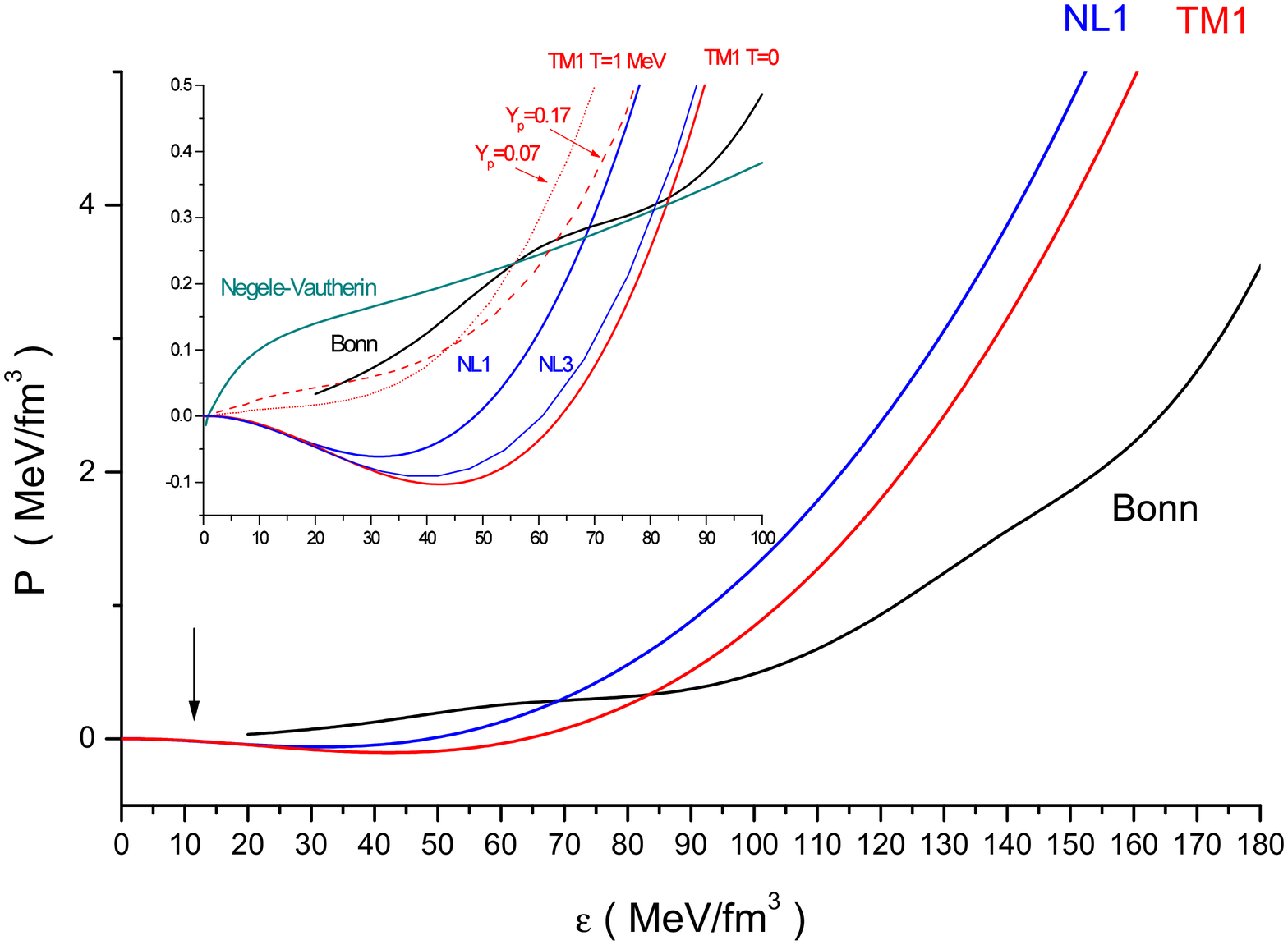}} \par}
\caption{\label{eos}The form of equation of state for the NL1, TM1 parameter sets together with the composite EOS. Variations of the
EOS with the proton fraction $Y_p$ are also presented.}
\end{figure}
\begin{figure}
{\par\centering \resizebox*{10cm}{!}{\includegraphics{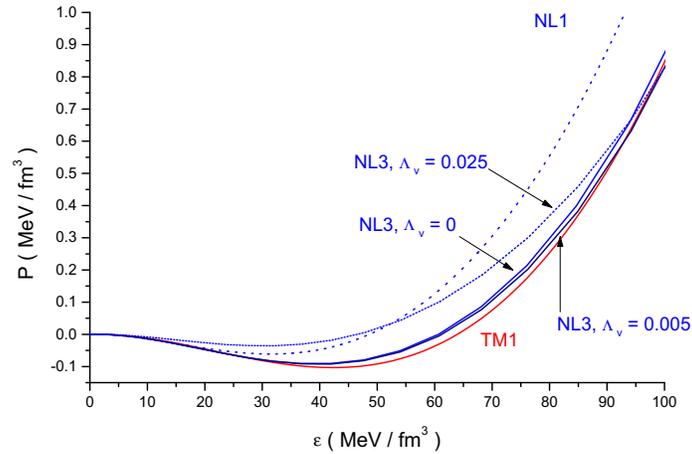}} \par}
\caption{\label{eosP}The equation of state for the original NL3 parameter set and for the NL3 supplemented with the $\Lambda_V$ coupling.}
\end{figure}
\begin{figure}
{\par\centering \resizebox*{10cm}{!}{\includegraphics{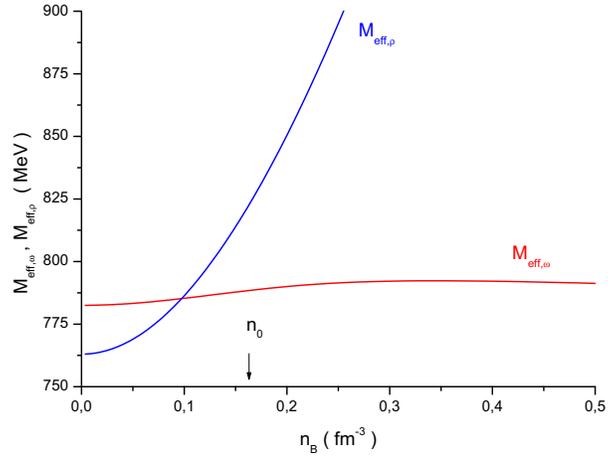}} \par}
\caption{\label{masy}The effective $\rho$ and $\omega$ meson masses.}
\end{figure}
\begin{figure}
{\par\centering \resizebox*{10cm}{!}{\includegraphics{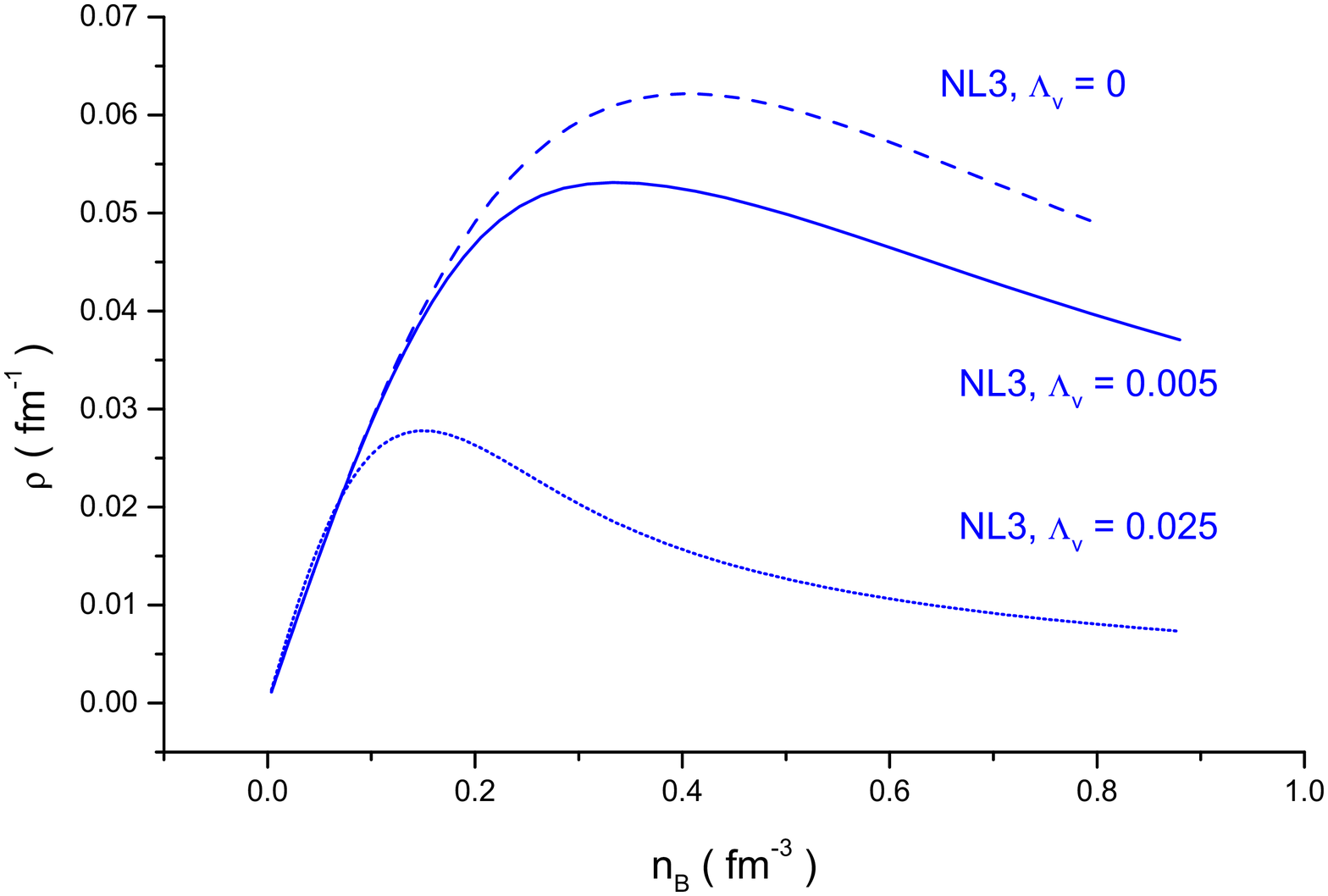}} \par}
\caption{\label{rho}The mean value of the $\rho$ meson field versus the baryon number density $n_B$.}
\end{figure}
\begin{figure}
{\par\centering \resizebox*{10cm}{!}{\includegraphics{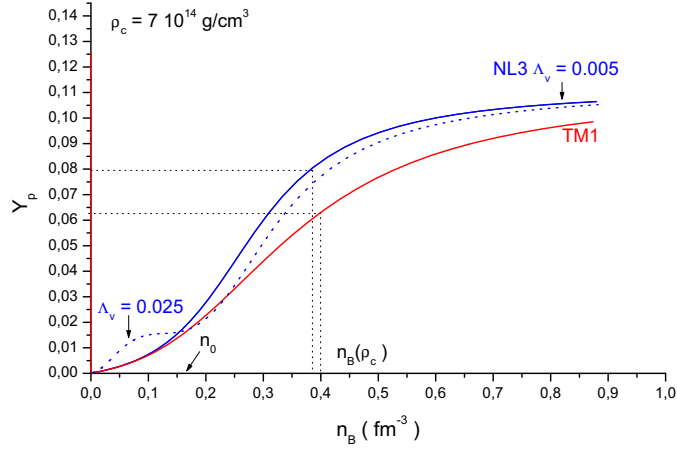}} \par}
\caption{\label{yp}The proton fraction versus the baryon number density $n_B$ for different parameter sets.}
\end{figure}
\newpage

\section{The geometry of the spacetime around the neutron star}
\vspace*{-0.5pt}
\noindent
The number of observed rotating neutron stars is increasing making it worthwhile
to study the influence of
rotation on the structure and parameters of neutron stars.
The spacetime
outside a rotating neutron star
is much more complicated than the metric outside a non-rotating
one.
It seems to be  interesting to investigate not only the properties of rotating neutron stars
but the phenomena occurring in the vicinity of the neutron stars as well.
In general the metric of a stationary, axisymmetric, asymptotically
flat spacetime has the form
\begin{eqnarray}
ds^2 & = & g_{\mu \nu}dx^{\mu }dx^{\nu } \\  \nonumber
     & =  & -e^{\gamma +\rho
}dt^2+e^{2\alpha }(dr^2+ r^2d\theta ^2)+r^2sin^2{\theta
}e^{(\gamma - \rho )}(d\varphi -\omega dt)^2
\end{eqnarray}
with $g_{\mu \nu}$ being the metric tensor. The metric potentials s $\gamma ,\rho $ and $\alpha $ and the
angular velocity $\omega $ of the stellar fluid are functions of the radial coordinate $r$ and the polar angle $\theta $ only.
The solution of Einstain equations for stationary rotating black holes, namely the Kerr metric
in Boyer-Lindquist coordinates and geometrized units $(G=c=1)$
is given by
\begin{eqnarray}
ds^2 & = & -(1-\frac{2Mr}{\Sigma})dt^2-(4Mar\frac{sin^2\theta }{\Sigma })dtd\varphi  \\ \nonumber
     & + & \frac{\Sigma }{
     \Delta }dr^2+
\Sigma d\theta  ^2+(r^2+a^2+2Ma^2r\frac{sin^2 \theta}{\Sigma})sin\theta d\varphi ^2
\end{eqnarray}
Here $ \Delta ,\Sigma $ and $a$ are of the following form
\begin{eqnarray}
\Delta & = & r^2-2Mr+a^2 \\ \nonumber
\Sigma & = & r^2+acos\theta  \\   \nonumber
a &=& \frac{J}{M}
\end{eqnarray}
where $J$ is the angular momentum.
Setting $a=0$ the above metric reduces to the Schwarzschild one
\begin{equation}
ds^2=-(1-\frac{2M}{r})dt^2+(1-\frac{2M}{r})^{-1}dr^2+r^2d\theta ^2+r^2sin^2\theta
d\phi^2.
\end{equation}
In this case the mass $M$ and the radius $R$ of the star are obtained by
numerical solution of the relativistic equations for hydrostatic
equilibrium, the Tolman-Oppenheimer-Volkoff equations (the TOV
equations). The RMF Lagrangian density function (\ref{lagrang}) now is supplemented by the
standard gravitational part ${\cal L}_G=R/2\kappa $.
In order to compute the structure of rapidly rotating fluid body
the numerical method developed by Butterworth and Ipser
\cite{12} is used.
Apart from this exact numerical treatment there is a perturbative
Hartle's method which is based on the assumption that rotating massive body is no longer spherically symmetric.
Both spherical and quadrupole deformations of the structure of the star are the effects of the rotation.
The star is destorted, and
expanding the metric functions through second order in the stars rotational velocity $\Omega $ one can obtain the following form
of the perturbed
metric
\begin{equation}
ds^2=-e^{\gamma +\rho }dt^2+e^{2\alpha }(r^2d\theta ^2+dr^2)+e^{\gamma -\rho }r^2sin^2\theta (d\phi -\omega dt^2)^2+O(\Omega ^3)
\end{equation}
where metric functions in this perturbed line element
can be calculated
from Einstein's field equations and given as solutions of Hartle's stellar structure equations,
$\omega $ is the same as in the nonperturbative line
element \cite{13}.
The metric functions are determined with the use of
Einstein equations
\begin{equation}
G_{\mu \nu }=8\pi T_{\mu \nu}.
\end{equation}
The matter source is assumed to be a perfect fluid with the
stress-energy tensor $T_{\mu \nu }$ given by
\begin{equation}
T_{\mu \nu}=(\varepsilon +P)u_{\mu}u_{\nu}+g_{\mu \nu}P
\end{equation}
where $\varepsilon $ is the total energy density, $P$ the pressure.
\vspace*{4pt}   
The assumption of uniform rotation of a star means that the value of rotational velocity $\Omega $
is constant throughout the star. For uniformly rotating bodies there is a relation among
components of the four-velocity vector $u^{\varphi }=\Omega u^{t}$.
The nonzero components of the four-velocity vector $u^{\mu}$ of the matter satisfying the normalization
condition $u^{\mu }u_{\mu }=-1$ are of the form
\begin{equation}
u^{t}=e^{-(\gamma +\rho )/2}(1-v^2)^{(-1/2)},\hspace{1cm}u^{\varphi }=\Omega u^{t}
\end{equation}
$v$ is the physical velocity of the fluid, relative to the local
zero angular momentum observer and can be expressed in terms of
the angular velocity $\Omega $  as
\begin{equation}
v=(\Omega -\omega )sin\theta e^{-\rho } \hspace{1cm} \Omega
=\frac{d\varphi }{dt}
\end{equation}
The absolute limit on
stable neutron star rotation is determined by the Kepler frequency $\nu _K$.
It determines the frequency at which the mass shedding at the
stellar equator sets in.
The result of the work of Haensel and Zdunik \cite{14} shows
that the value of the Kepler frequency can be estimated knowing
the value of the mass and radius of the corresponding nonrotating
star and the empirical relation was given
\begin{equation}
\Omega_{K} \approx
C_{Hz}\sqrt{(M_s/M_{\odot})(R_s/10km)^3}=(0.63-0.67)\times
\Omega_{c}
\end{equation}
where, $C_{Hz}=7700 s^{-1}$ and $\Omega_{c}$ is the Newtonian value and is equal
\begin{equation}
\Omega_{c}=\sqrt{M_s/R_s^3}
\end{equation}
the index $s$ indicates that these values refer to the spherical
configuration. From the condition of stationarity and axisymmetry
of the metric it is clear that the energy and angular momentum
are constants of motion. One of the problem considered in this
paper is connected with the existence and location of the
marginally stable orbit in the equatorial plane outside the star.
In agreement with the predictions of the general theory of
relativity for  sufficiently compact stars there is a region
around the star, in the equatorial plane in which particles
moving along geodesics are unstable to radial perturbations.
Using the method presented by Bardeen \cite{15}
 one can calculate the location of the marginally stable
orbit. From the condition $p_{\mu }p^{\mu }=-1$ the equations of
motion of a particle confined to the equatorial plane have the
following form
\begin{equation}
\frac{dt}{ds}=p^{0}=e^{-(\gamma +\rho )}(E-\omega \Phi)
\end{equation}
\begin{equation}
\frac{d\Phi }{ds}=p^{1}=\Omega p^{0}=\frac{\Phi}{r^{2}e^{\gamma -\rho }}
+\omega e^{-(\gamma +\rho )}(E-\omega \Phi )
\end{equation}
\begin{equation}
e^{(2\alpha +\gamma +\rho )}(\frac{dr}{ds})^2 = Q(r,E,\phi )=e^{-(\gamma +\rho)}(E-\Phi \omega )^2-1-\frac{\Phi ^2}{r^2}e^{-\gamma +\rho}
\end{equation}
where  the constants of  motion are $E=-p_{0}$, the energy per unit mass, and $\Phi
=p_{1}$,
the angular momentum per unit mass about the axis of symmetry, $s$ is the proper time
along the geodesic.
The condition for the circular orbit is given by
\begin{equation}
\frac{\partial Q}{\partial r}=0
\end{equation}
This allows us to determine the velocity of rotation of a circular orbit
\begin{equation}
\upsilon =\pm (\sqrt{e^{-2\rho}r^2\omega ^{'2}+2r(\gamma ^{'}+\rho ^{'})+r^2(\gamma ^{'2}-\rho ^{'2})\pm e^{-\rho}r^2\omega ^{'}})/(2+r(\gamma ^{'}-\rho ^{'}))
\end{equation}
Solutions of this equation determine the velocity of co-rotate and counter-rotate particles respectively,
whereas the condition for the stability of the circular orbit is fulfilled  when
\begin{equation}
\frac{\partial ^2Q}{\partial r^2}< 0
\end{equation}
Location of the innermost marginally stable orbit is given by the
requirement
\begin{equation}
\frac{\partial ^2Q}{\partial r^2}=0
\end{equation}
which leads to the following equation
\begin{eqnarray}
(1-v^2)\frac{\partial V}{\partial r^2} & = & -(\gamma ^{''}+\rho ^{''})(1+v^2)-v^2(-\gamma ^{'}+\rho^{'})((-\gamma ^{'}+\rho ^{'})+2)   \\ \nonumber
                                       & + & (\gamma ^{'}+\rho ^{'})^2 +4\frac{v^2\omega ^{'}}{(\Omega -\omega)}(\gamma ^{'}+\rho ^{'})-2\frac{v^2\omega ^{''}}{(\Omega -\omega)} \\ \nonumber
                                       & - & 6\frac{v^2}{r^2}+2\frac{v^2}{r}(-\gamma ^{'}+\rho ^{'})+2\frac{v^4\omega ^{'2}}{(\Omega -\omega)^2}
\end{eqnarray}
where  prime denotes a first order partial derivative with respect to $r$.
This allows to obtain the equation for the  radius
of the marginally stable orbit.
Considering the case of a nonrotating, spherically symmetric star the mass of the neutron
star as a function of the marginally stable orbit radius can be
expressed as $M=R_{ms}/6$ and the orbital frequency of a point
particle in a circular orbit is given by
\begin{equation}
\nu  =\frac{1}{2\pi}\sqrt{\frac{M}{r^3}}
 \label{orbit}
\end{equation}
$M$ is the neutron star mass, $r$ is the orbit radius.
The $M=R_{ms}/6$ relation on the mass- radius plane is a straight line which
intersects the function $M(R_{orb},\nu)$ (\ref{orbit}) at the point
\begin{equation}
M_{max}=\frac{2200Hz}{\nu}M_{\odot}.
\end{equation}
$M_{max}$ settles the upper bound on the mass of the star.
In the case of rotating neutron stars the obtained mass-orbit
radius relation is altered. The results depend on the rate of
stellar spin which can be characterized by the dimensionless
parameter $j=J/M^2$, $J$ and $M$ are the angular momentum and the
mass of the star. For very small values of $j$ the approximate
expressions valid to first order of the parameter $j$ can be
incorporated in order to achieve the expressions for the
frequency of gas in the circular orbit $\nu$, the
Lense-Thirring precession frequency $\nu _{LT}=\omega /(2\pi )$
and adequate expressions for the marginally stable orbit
\begin{equation}
\nu =\frac{1}{2\pi}\sqrt{\frac{M}{r^3}}(1-(\frac{M}{r})^{\frac{3}{2}}j)
\end{equation}
\begin{equation}
\nu_{LT}=\frac{2}{2\pi}\frac{M^2}{r^3}j
\end{equation}
\begin{equation}
r_{ms}=6M(1-(\frac{2}{3})^{\frac{3}{2}}j)
\end{equation}
\begin{equation}
\nu _{ms}=\frac{6^{-3/2}}{2\pi M}(1+\frac{11}{6^{3/2}}j)
\end{equation}
\begin{equation}
\nu_{LT,ms}=\frac{1}{6^3\pi}\frac{j}{M}
\end{equation}
where $r_{ms}$ is the radius of the marginally stable orbit for
the rotating neutron star.
\begin{figure}
{\par\centering \resizebox*{10cm}{!}{\includegraphics{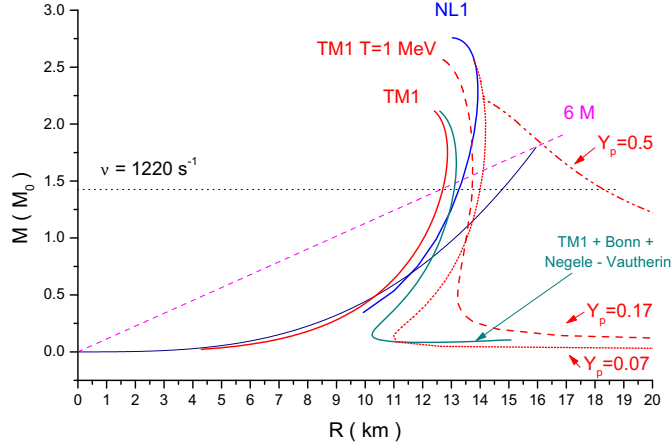}} \par}
\caption{\label{margrm}The mass-radius relations
for the employed equations of state.
The neutron star mass as a function of the marginally stable orbit radius is shown (the dotted line) together with the
relation $M(R_{orb},\nu)$}
\end{figure}
\begin{figure}
{\par\centering \resizebox*{10cm}{!}{\includegraphics{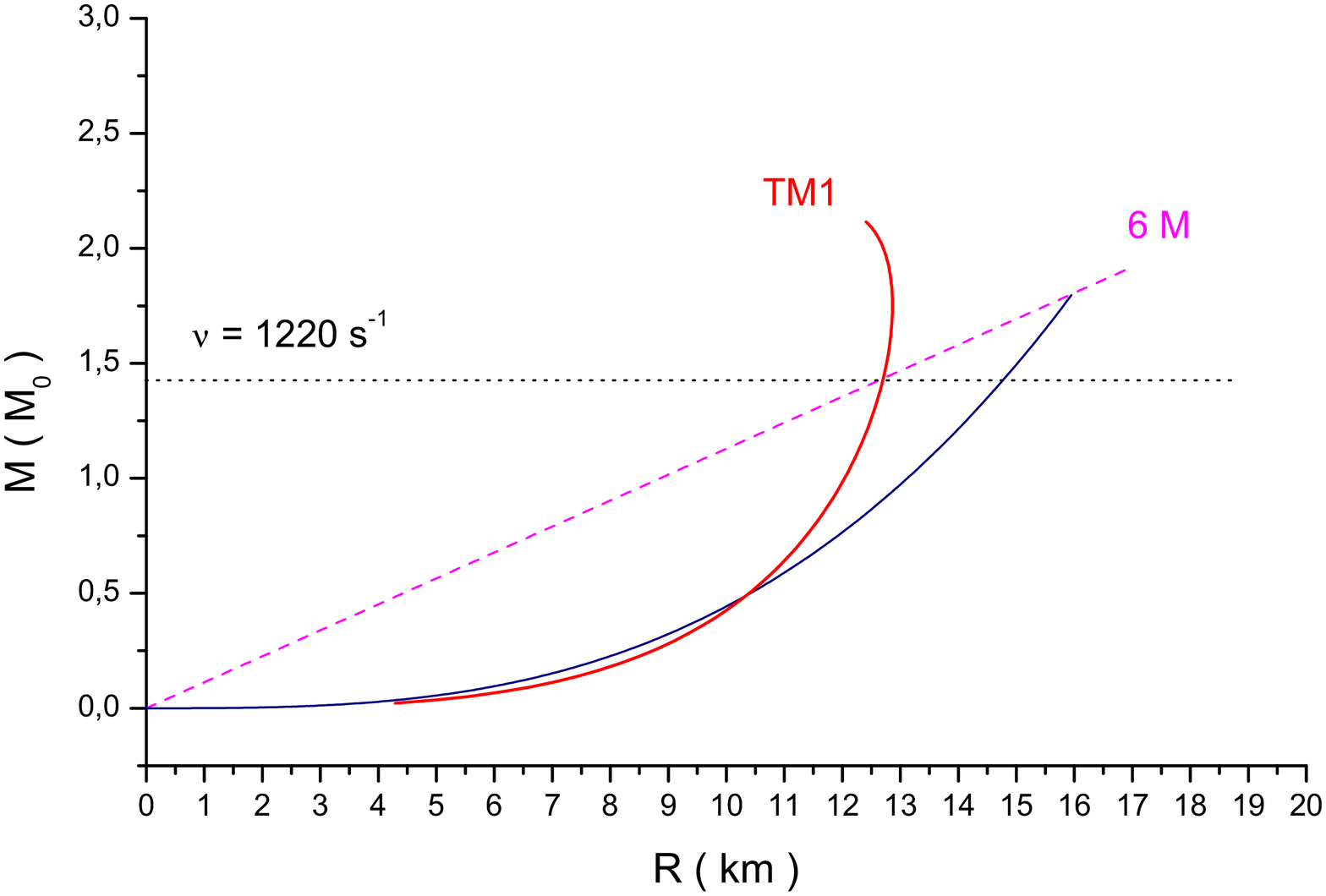}} \par}
\caption{\label{margrm1}The mass-radius relation for the TM1 parameter set with the relation $M(R_{orb},\nu)$. Dotted line represents
the  of neutron star mass as a function of the marginally stable orbit radius}
\end{figure}
\newline
\section{Conclusions}
\vspace*{-0.5pt}
\noindent

The properties of non-rotating neutron stars for given equations
of state are listed in Table 2. They  correspond to the maximum
stable star configuration. The employed parameter sets are the
TM1 and NL1 for the neutron rich matter without crust and
$TM1^{*}$ which marks the composite EOS namely TM1 one joined with
those of Negele and Vautherin and Haensel and Pichon. For the latter case
the numerical calculations were performed for the density span
from $7 \times 10^{14}g/cm^3$ (the maximum density for the TM1
parameter set) to $3\times 10^7 g/cm^3$. In the obtained
equations of state the original NL3 parameter set and the NL3 one
supplemented with the additional $\rho -\omega$ mesons coupling
have been also analyzed.
For comparison the neutron star parameters for TM1 EOS with changed proton fraction have been obtained.
The forms of the equations of state are
presented in Figs. \ref{eos} and \ref{eosP}.
\begin{figure}
{\par\centering
\resizebox*{10cm}{!}{\includegraphics{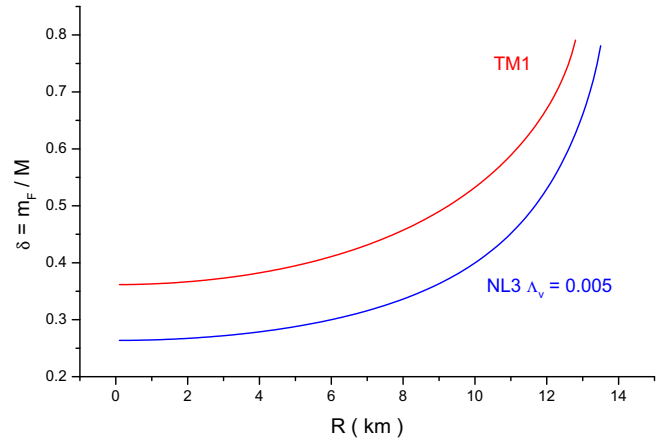}}
\par} \caption{\label{delta} The effective nucleon mass versus the neutron star radius for the neutron-rich neutron star interior
for TM1 and NL3 ($\Lambda_V=0.005$) parameter sets.}
\end{figure}
\begin{figure}
{\par\centering \resizebox*{10cm}{!}{\includegraphics{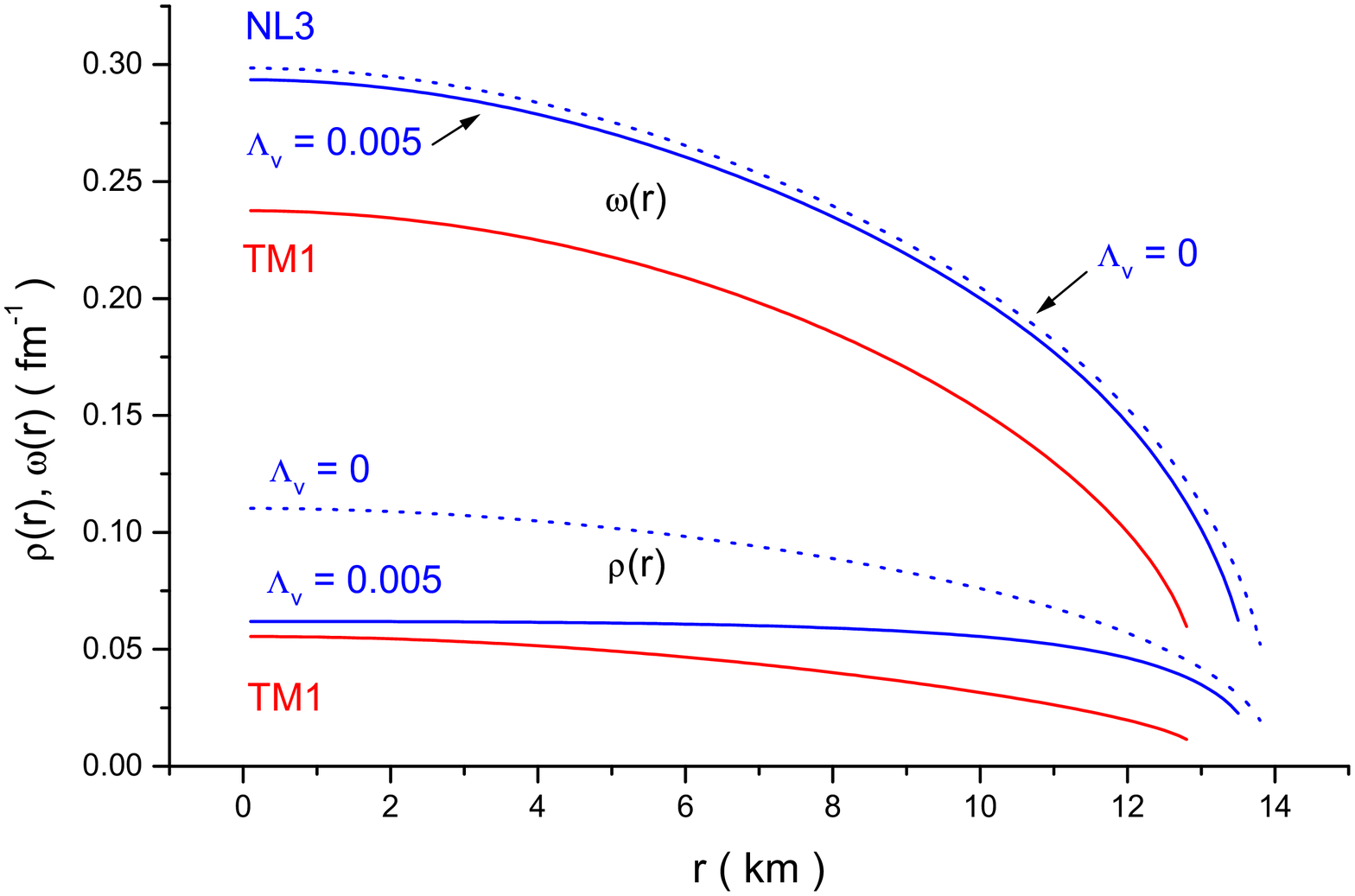}} \par}
\caption{\label{rhost}The $\rho$ and $\omega$ field as a function of stellar radius for different parameter sets.}
\end{figure}
\begin{figure}
{\par\centering \resizebox*{10cm}{!}{\includegraphics{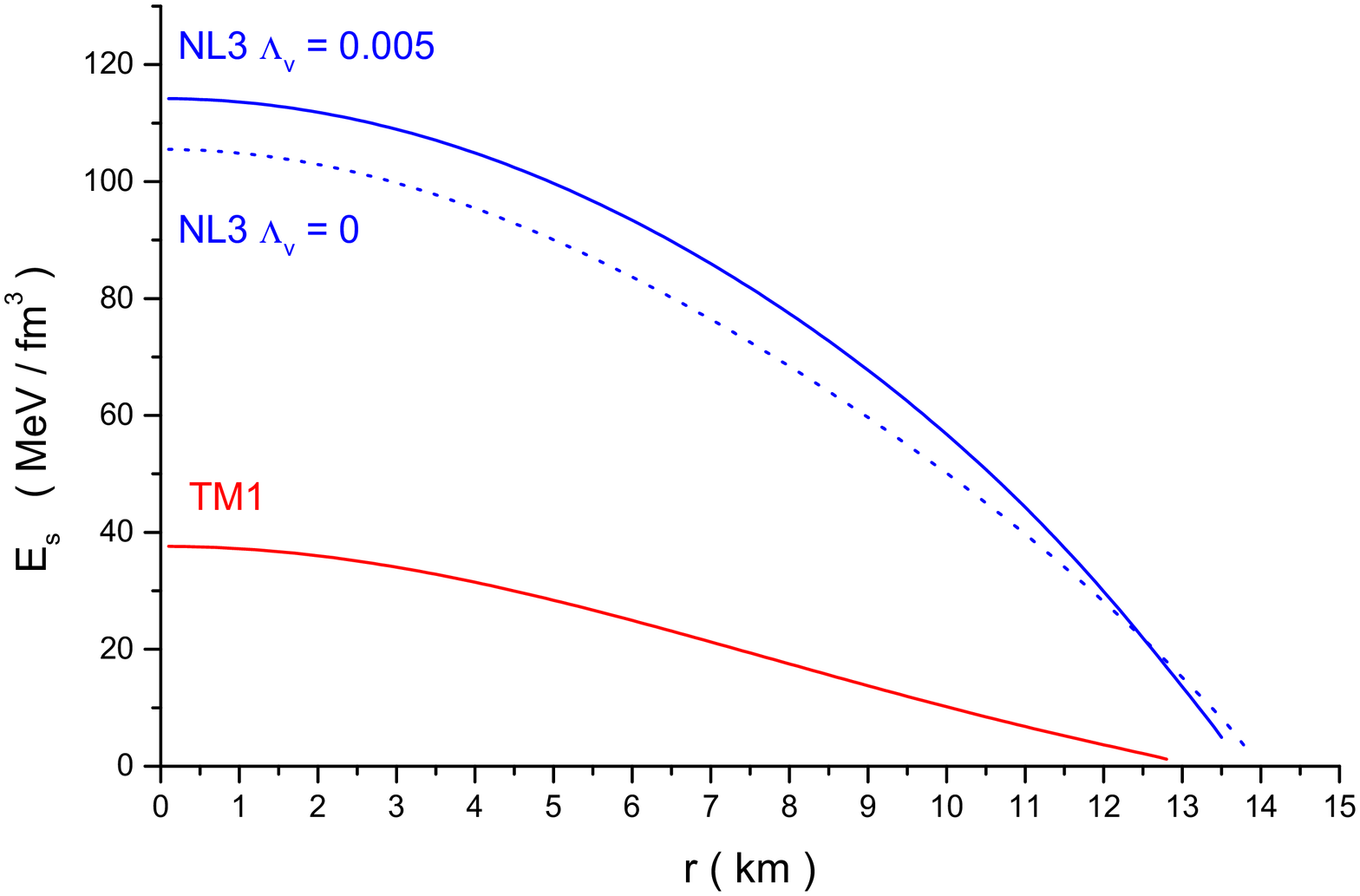}} \par}
\caption{\label{engs} The energy symmetry versus the stellar radius for TM1 and the original NL3 ($\Lambda_V=0$) and
the extended NL3 ($\Lambda_V=0.005$) models.}
\end{figure}

Each particular column of Table 2 contains: the
central density $\rho _c$, the baryon rest mass $M_B$, the
gravitational mass $M$, the radius $R_s$, the moment of inertia
$I$, the value of the Newtonian limit of the stellar spin
$\Omega_{c}$
and the gravitational redshift $z$.

\begin{table}[htbp]
\tcaption{\label{tab3}Calculated properties of nonrotating (limiting-mass for TM1) neutron star models
with \protect\protect\( \rho _{c}=7\, 10^{14}\, g/cm^{3}\protect \protect \)
for TM1, NL1, for TM1{*} (TM1+Bonn \cite{4}+Negele,Vautherin )
and generalized NL3 (\protect\( \Lambda _{v}=0.005\protect \)) \cite{3}
parameter sets.}
\centerline{\footnotesize\smalllineskip
\begin{tabular}{l c c c c c c }
\hline
EOS&
 \( R_{s}\, (km) \)&
 \( M\, (M_{\odot }) \)&
 \( M_{B}\, (M_{\odot }) \)&
 \( I\, (M_{\odot }km^{2}) \)&
 \( \Omega _{c}\, (10^{4}\, s^{-1}) \)&
 \( z \)\\
\hline
TM1&
 \( 12.866 \)&
 \( 1.643 \)&
 \( 1.978 \)&
 \( 113.89 \)&
 \( 1.0112 \)&
 \( 0.26715 \)\\
\hline
NL1&
 \( 13.971 \)&
 \( 2.231 \)&
 \( 2.935 \)&
 \( 196.32 \)&
 \( 1.0422 \)&
 \( 0.37588 \)\\
\hline
TM1{*}&
 \( 14.644 \)&
 \( 1.644 \)&
 \( 1.982 \)&
 \( 114.37 \)&
 \( 0.8338 \)&
 \( 0.2234 \)\\
\hline
TM1 (\( Y_{p}=0.07 \))&
 \( 14.029 \)&
 \( 1.915 \)&
 \( 2.279 \)&
 \( 154.04 \)&
 \( 0.9593 \)&
 \( 0.29442 \) \\
\hline
NL3 (\( \Lambda _{v}=0.005 \))&
\( 13.573 \)&
\( 2.105 \)&
\( 2.742 \)&
\( 174.84 \)&
\( 1.0570 \)&
\( 0.3584 \)\\
\hline
\end{tabular}}
\end{table}
\begin{table}[htbp]
\tcaption{\label{tab4}
    Calculated properties of slowly rotating (limiting-mass for TM1) neutron
    star models with \protect\protect\(
    \rho _{c}=7\, 10^{14}\, g/cm^{3}\protect \protect \)
    and \protect\protect\( \Omega =10^{3}\, s^{-1}\protect \protect \).}
\centerline{\footnotesize\smalllineskip
\begin{tabular}{l c c c c c c c }
\hline
EOS&
 \( R_{s} \)&
 \( R_{e} \)&
 \( M_{f} \)&
 \( R_{+} \)(\( 6\, M_{f} \))&
 \( R_{-} \)&
 \( v_{K}\, (Hz) \)&
 \( v_{LT}\, (Hz) \)\\
\hline
TM1&
 \( 12.881 \)&
 \( 12.904 \)&
 \( 1.648 \)&
 \( 15.304 \) (\( 14.57 \))&
 \( 13.794 \)&
 \( 2228.5 \)&
 \( 111.6 \)\\
\hline
NL1&
 \( 13.982 \)&
 \( 13.996 \)&
 \( 2.238 \)&
 \( 20.718 \) (\( 19.78 \))&
 \( 18.601 \)&
 \( 1551.6 \)&
 \( 58.1 \)\\
\hline
TM1{*}&
 \( 14.667 \)&
 \( 14.695 \)&
 \( 1.652 \)&
 \( 15.323 \) (\( 14.58 \))&
 \( 10.514 \)&
 \( 2114.0 \)&
 \( 62.2 \)\\
\hline
TM1  (\( Y_{p}=0.07 \))&
 \( 14.046 \)&
 \( 14.064 \)&
 \( 1.928 \)&
 \( 17.823 \) (\( 16.96 \))&
 \( 16.080 \)&
 \( 1815.8 \)&
 \( 61.8 \) \\
\hline
NL3  (\( \Lambda _{v}=0.005 \))&
\( 13.585 \)&
\( 13.598 \)&
\( 2.110 \)&
\( 19.547 \) (\( 18.66 \))&
\( 17.735 \)&
\( 1644.9 \)&
\( 58.1 \)\\
\hline
\end{tabular}}
\end{table}

The observed X-ray radiation can be used to investigate phenomena occuring in the strong gravitational fields near neutron stars.
With the Rossi X-Ray Timing Explorer  high-frequency X-ray brightness oscillations from several low-mass X-ray binaries
have been discovered.
Observations of kilohertz QPO sources can show the existence of circular orbits around neutron stars.
The higher frequency
in a kilohertz pair in the sonic point model  \cite{16} can be explained as the orbital frequency of gas in a
nearly circular orbit around the neutron star which
in the case of Schwarzschild metric  is given by the relation
(\ref{orbit}.)
From this equation
one can obtain the relation between the orbit radius $r$ and the mass of
the neutron star for given value of the frequency $\nu _{QPO}$.
Thus the orbit radius is determined by the mass of the neutron star which in turn depends on the particular choice of the equation of state,
and as a consequence the allowed region on the mass-radius plane for the
observed values of $\nu _{QPO}$ exists.
Fig. 6 shows the mass-radius plane together with the allowed area for the fixed value of the frequency $\nu _{QPO}$  in the case of non-rotating stars.
Having evaluated the EOS the numerical solutions of the structure
equations are obtained
and then the mass-radius relations are constructed for chosen forms of the equations of state listed in Table 1.
Table 3 contains properties of slowly rotating neutron stars for mentioned earlier equations of state. $R_{+}$ and $R_{-}$ denote
radii of marginally stable orbits for co-rotating and counter-rotating particles respectively.
The condition $R_{s} < R_{ms}$ (the neutron star radius $R_{s}$ is  smaller than the radius of the marginally stable orbit $R_{ms}$)
is fulfilled for the employed equations of state.
For the TM1 parameter set there exist masses in the relevant range from $0.5 M_{\odot}$ to $1.4 M_{\odot}$
for which the condition $R_{s}< R_{orb}$ is fulfilled, for LN1 the same range is from $1.2 M_{\odot}$ to $1.6 M_{\odot}$.
The composite EOS changes the range from $0.75 M_{\odot}$ to $1.5 M_{\odot}$.
For the composite form of the equations of state the minimal star configuration appears.
The same result is achieved changing the proton fraction $Y_{p}$.
From Fig.6 one can see that the minimum mass configuration is very sensitive to the proton fraction.
The symmetric nuclear matter excluded the existence of stable minimum configuration stars.
The  value of the proton fraction $Y_{p}$ is directly connected with the energy symmetry $E_{s}$, hence the
influence of the symmetry energy on neutron star parameters, especially on the minimum configuration is straightforward
and comes from the change of the chemical composition of the neutron star matter.
Fig.10 compares the symmetry energy $E_{s}$ as a function of stellar radius $R$ for parameter sets with (NL3 $\Lambda_{V}=0.005$)
and without (TM1 and NL3 $\Lambda _{V}=0$) additional meson coupling. In the first case the  symmetry energy prevails over
the latter ones. The changes in energy symmetry are caused by the appearance of the effective meson masses which are closely connected
with the self-consistency equation for meson fields. Using these equations the form of the meson fields expectation values as function
 of stellar radius $R$ for different parameter sets are presented
in Fig.9. In the case of different from zero $\Lambda_{V}$ the value of $\rho$ decreases and the energy symmetry tends to the
value characteristic to the symmetric matter. This is of special significance  for the  minimum configuration models. Solutions with minimal masses are
sensitive to the proton fraction which in turn is determined by the value of parameter $\Lambda_{V}$. Thus,  the minimum mass configurations
are sensitive to the existence of the additional meson coupling and one can conclude that the nonzero $\Lambda_{V}$ modies the mass radius dependence.
The effective nucleon mass $M^*$ is caused by the nucleon-scalar field coupling. The parameter $\delta =\frac{m_F}{M}$ versus the stellar
radius $R$ is presented in Fig.8 for TM1 and NL3 $\Lambda_{V}=0.005$.
The idea of introducing the additional meson coupling and finding an analogy between the neutron
rich nucleus ${}^{209}Pb$ and the behaviour
of neutron star crust was introduced by Horowitz and Piekarewicz in their work \cite{2}.
As it has been stated in the previous chapter the change of proton fraction $Y_{p}$ and the behaviour of the energy symmetry are
connected to the presence of the additional $\rho-\omega $ coupling and the nonzero value of $\Lambda _{V}$.
The next step is the extension of EOS for neutron rich matter to
the case of finite temperature \cite{17}. This will be the subject of future investigations.
Since the neutron star parameters varies with temperature, which
the allowed area on the mass-radius plane changes as well.
For temperatures different from zero the mass and the radius of
the star have higher values and any
configuration with larger mass altered a
marginally
stable orbit radius.
The most sensitive component of neutron star to temperature are electrons (in the case when neutrinos are neglected), thus the
influence of temperature acts in  the similar  way as the introduction of the parameter $\Lambda_{V}$.

\end{document}